\documentclass[main,dvipsnames]{siamart220329}

\usepackage{amsfonts,subfigure}
\usepackage{graphicx}
\usepackage{epstopdf}
\usepackage{algorithmic}
\usepackage{hyperref}
 \hypersetup{
     colorlinks=true,
     linkcolor=black,
     filecolor=black,
     citecolor = black,      
     urlcolor=black,
     }
\usepackage{amsmath, amssymb}
\usepackage{relsize}
\usepackage{pifont}
\usepackage{enumitem}
\usepackage{oubraces}
\usepackage{listings}
\usepackage{booktabs}
\usepackage{xfrac}
\usepackage{cleveref}
\usepackage{caption}
\usepackage{tabularx}
\usepackage{array}
\usepackage{multirow}
\usepackage{mathrsfs}
\usepackage{microtype}
\usepackage{natbib}
    
\usepackage{tikz}
\usetikzlibrary{arrows,backgrounds,patterns}
\usetikzlibrary{matrix,shapes,fit,calc,shadows,plotmarks}
\usetikzlibrary{decorations}
\usetikzlibrary{decorations.markings}
\RequirePackage{pgfplotstable}
\RequirePackage{pgfplots}
\usepackage{pgfplotstable}
\usepackage{xparse}
\usepackage[T1]{fontenc}
\usepackage[utf8]{inputenc}
\usepackage[normalem]{ulem}
\usepackage{xcolor}
\usepackage{todonotes}

\newcommand{\la}{\lambda}
\newcommand{\laexp}{\lambda_{\mathrm{exp}}}
\newcommand{\ra}{\rightarrow}

\newcommand\eps{\epsilon}
\renewcommand*{\i}{\mathrm{i}}
\renewcommand*{\d}{\mathrm{d}}
\newcommand*{\e}{\mathrm{e}}

\newcommand*{\ep}{\epsilon}
\newcommand*{\de}{\operatorname{d\!}{}} 
\newcommand{\sdd}[2]{\frac{\de^2#1}{\de#2^2}}
\newcommand{\dd}[2]{\frac{\de#1}{\de#2}}

\def\Xint#1{\mathchoice
   {\XXint\displaystyle\textstyle{#1}}%
   {\XXint\textstyle\scriptstyle{#1}}%
   {\XXint\scriptstyle\scriptscriptstyle{#1}}%
   {\XXint\scriptscriptstyle\scriptscriptstyle{#1}}%
   \!\int}
\def\XXint#1#2#3{{\setbox0=\hbox{$#1{#2#3}{\int}$}
     \vcenter{\hbox{$#2#3$}}\kern-.5\wd0}}

\author{Josh Shelton\thanks{Department of Mathematical Sciences, University of Bath, BA2 7AY, UK} 
\and Philippe H. Trinh
\and S. Jonathan Chapman\thanks{Oxford Centre for Industrial and Applied Mathematics, Mathematical Institute, \newline University of Oxford, OX1 3LB, UK} 
}

\title{Pathological exponential asymptotics for a model problem of an equatorially trapped Rossby wave
\thanks{Submitted to the editors 09/02/2023.}
\funding{PHT is supported by the Engineering and Physical Sciences Research Council [EP/V012479/1].}
}

\author{Josh Shelton\thanks{Department of Mathematical Sciences, University of Bath, BA2 7AY, UK} 
\and S. Jonathan Chapman\thanks{Oxford Centre for Industrial and Applied Mathematics, Mathematical Institute, \newline University of Oxford, OX1 3LB, UK} 
\and Philippe H. Trinh\footnotemark[2]
}

\begin{document}

\maketitle
\begin{abstract}
We examine a misleadingly simple linear second-order eigenvalue problem (the Hermite-with-pole equation) that was previously proposed as a model problem of an equatorially-trapped Rossby wave. In the singularly perturbed limit representing small latitudinal shear, the eigenvalue contains an exponentially-small imaginary part; the derivation of this component requires exponential asymptotics. In this work, we demonstrate that the problem contains a number of pathological elements in exponential asymptotics that were not remarked upon in the original studies. This includes the presence of dominant divergent eigenvalues, non-standard divergence of the eigenfunctions, and inactive Stokes lines due to the higher-order Stokes phenomenon. The techniques developed in this work can be generalised to other linear or nonlinear eigenvalue problems involving asymptotics beyond-all-orders where such pathologies are present.
\end{abstract}


\pagestyle{myheadings}
\thispagestyle{plain}
\markboth{Shelton, Chapman, and Trinh}{The Hermite-with-pole equation}

\section{Introduction} 
The motivation of this work stems from an interesting mathematical model that was proposed by \cite{boyd1998sturm} in order to describe equatorially-trapped Rossby waves when the mean shear flow is only a function of the latitude. In such cases, the eigenfunctions are modelled by the so-called \textit{Hermite-with-pole} equation 
\begin{subequations}
\begin{gather} \label{eq:originalu}
\dd{^2 u}{z^2} + \left[\frac{1}{z} - \lambda - \left(z - \frac{1}{\ep}\right)^2\right]u = 0, \\
u(z) \to 0 \quad \text{as} \quad z \to \pm \infty, \\
u(0)=1.
\end{gather}
\end{subequations}
Here, $\epsilon$ corresponds to the shear strength, $u$ corresponds to a normal mode amplitude, and $\lambda$ is an eigenvalue determined by the boundary condition at $z=0$. Although this resembles the standard parabolic cylinder equation with Hermite functions as eigenfunctions, the pole at $z = 0$ lies in the interval of consideration. Boyd \& Natarov consider the pole at $z = 0$ as emerging from a singularity in the analytic continuation, which approaches the real axis as viscosity tends to zero. As it turns out, the associated eigenvalue to \eqref{eq:originalu} is complex-valued; in the limit $\ep \to 0$, the eigenvalue contains an exponentially-small imaginary part, $\text{Im} [ \lambda ] = O(\e^{-1/\ep^2})$. One of the aims of the analysis is to derive this exponentially-small eigenvalue component.

 \begin{figure}[htpb]
	\centering
	\includegraphics[]{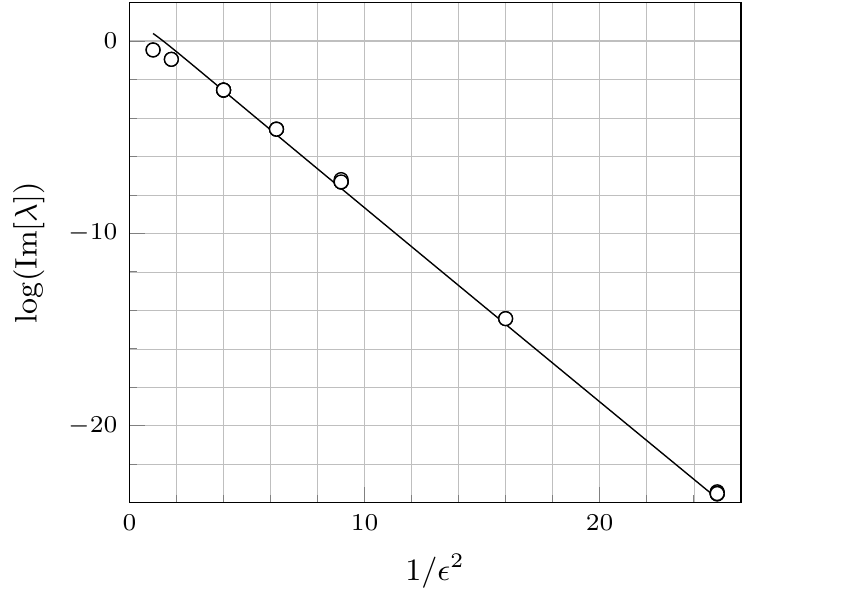}
	\caption{The imaginary component of the eigenvalue, $\lambda$,
          is shown for the numerical solutions of \cite{boyd1998sturm}
          (circles) and the analytical prediction of
          \mbox{$\text{Im}[\lambda]=\sqrt{\pi}[1-2\ep \log
            \ep+\ep(\log 2+\gamma)]\e^{-1/\ep^2}$} (line). Here,
          $\gamma \approx 0.577$ is the Euler-Macheroni constant.}%
	\label{fig:ImEig}
\end{figure}

In their work, \cite{boyd1998sturm} note that an asymptotic expansion of $u(z)$ in integer powers of $\epsilon$ diverges, and they develop a procedure for approximating $\text{Im} [\lambda]$ with the use of an integral property from Sturm-Louville theory. Their approach relies upon the use of special functions theory and the niceties of the linear differential equation. In contrast, the emphasis of our work here will be on developing a framework that is applicable for more general differential equations---particularly for nonlinear problems where special functions theory is unavailable. Our goal is to study the divergence of the asymptotic expansion for the eigenfunction and examine its connection, via the Stokes phenomenon, to the exponentially-small components. In \S\ref{sec:conclusion} we discuss the significance in the application of techniques developed in this paper, both to the more complete geophysical problem discussed by \cite{natarov2001beyond}, as well as other problems involving singular perturbations.

For analysis, there is a more convenient form of \eqref{eq:originalu}, which is found by shifting 
\begin{gather}
y = z - \frac{1}{\ep},
\end{gather}
where now $y = 0$ corresponds to the equator. Then we have, for $u = u(y)$, 
\begin{equation}
\dd{^2 u}{y^2} + \left[ \frac{\ep}{1 + \ep y} - y^2 \right]u = \lambda u.
\end{equation}
This intermediary equation contains a turning point at $y =-1/\ep$ which we study by rescaling with $y=Y/\ep$.
We then set $u(y) = \e^{-y^2/2} \psi(Y)$ which yields the system \begin{subequations}
\begin{gather}
\ep^2 \psi^{\prime \prime} - 2Y \psi ^{\prime} + \frac{\ep \psi}{1+Y}= (\lambda +1) \psi, \label{eq:hermite} \\
\e^{-{Y^2}/{2\ep^2}} \psi(Y) \to 0 \quad \text{as} \quad  Y \to \pm\infty, \label{eq:bc1} \\
\psi(0)=1. \label{eq:bc2} 
\end{gather}
\end{subequations}
In \eqref{eq:hermite} and henceforth, we use primes ($^\prime$) to denote differentiation in $Y$.

\section{A roadmap of the methodology and main results}\label{sec:roadmap}
As it turns out, the Hermite-with-pole problem \eqref{eq:hermite} has a number of non-trivial elements that were not remarked upon in the original studies; the treatment of which has required the development of new techniques in exponential asymptotics. We explain some of these aspects in the context of singularly perturbed linear eigenvalue problems of the form \eqref{eq:hermite}, $\mathcal{L}(\psi; \ep) = \lambda \psi$, although many of the same ideas apply more generally to nonlinear eigenvalue problems. 

Firstly, asymptotic expansions for $\psi = \psi_0 + \ep \psi_1 + \cdots$ and $\lambda = \lambda_0 + \ep \lambda_1 + \cdots$ are sought, but these expansions are divergent and must be optimally truncated. The solution is then expressed as a truncated series with a remainder by considering
\begin{equation}
\label{eq:remainder}
\psi(Y) = \sum_{n=0}^{N-1} \ep^n \psi_n(Y)+\mathcal{R}_N(Y),
\end{equation}
with a similar expression for the eigenvalue, $\lambda$. When $N$ is chosen optimally [later shown to be of $O(\ep^{-2})$] the remainder $\mathcal{R}_N(Y)$ is exponentially-small, and satisfies the linear eigenvalue problem of
\begin{equation}
\label{eq:remainderEQ}
\mathcal{L}(\mathcal{R_N};\ep) \sim -\ep^N \psi_{N-2}^{\prime \prime}.
\end{equation}
The remainder, $\mathcal{R}_N$, will exhibit the Stokes phenomenon, in which its magnitude rapidly varies across certain contours in the complex $Y$-plane. Indeed, as we shall show, this behaviour can be predicted by estimating the growth of the forcing term $\psi_{N-2}^{\prime \prime}$. Thus, the late-term behaviour of the divergent series, $\psi_N$ with $N \to \infty$, is required in order to correctly resolve the Stokes phenomenon on the remainder $\mathcal{R}_N(Y)$. This `decoding' of divergence is one of the hallmarks of exponential asymptotics.

One of our main results of this paper is that for the Hermite-with-pole problem, additional components of the late-term divergence, $\psi_n$, are required. It is well known, according to the principles of exponential asymptotics (cf. \cite{chapman_2002}) that the $n$th-order approximation of most singularly perturbed differential equations exhibits a factorial-power-divergence similar to
\begin{equation}\label{eq:typicalfactorial}
    \psi_n \sim \frac{Q(Y)\Gamma\left(\frac{n}{2}+\alpha\right)}{\chi(Y)^{\frac{n}{2}+\alpha}} \quad \text{as $n\to \infty$},
\end{equation}
where different problems may involve slight modifications to the above form. Thus for instance, the fractional coefficient of $n$ that appears above may be modified to ensure the correct dominant balance arises in the equation. The functions $Q$ and $\chi$ and the constant $\alpha$ prescribe the divergent behaviour.

However in this work we demonstrate that the Hermite-with-pole problem
exhibits an atypical divergence of the form
\begin{equation} \label{eq:LateSolIntro}
\psi_{n} \sim
\left\{\begin{aligned}
 &~\mathcal{S}(Y) \Big[ L(Y)\log{(n)}+Q(Y) \Big]\frac{\Gamma(\frac{n}{2} +\alpha_0)}{\chi^{n/2 +\alpha_0}}\qquad \\
 & \qquad \qquad \qquad \qquad \quad +Q^{(\lambda_n)}_{0}(Y)\log^2{(n)}\Gamma\bigg(\frac{n+1}{2}+\alpha_0\bigg) \quad \text{for $n$ even}, \\
 &\underbrace{\mathcal{S}(Y)}_{\text{HOSP}}\underbrace{R(Y)\frac{\Gamma(\frac{n}{2}+\alpha_1)}{\chi^{n/2+\alpha_1}}}_{\text{na\"ive divergence}} +\underbrace{R^{(\lambda_n)}_{1}(Y)\log{(n)}\Gamma\bigg(\frac{n+1}{2}+\alpha_1\bigg)}_{{\lambda_n \text{ divergence}}} \quad \text{for $n$ odd}.
\end{aligned}\right.
\end{equation}
Here, the singulant, $\chi(Y)$, takes a value of zero at singularities in the early orders of the asymptotic expansion, and $\mathcal{S}(Y)$ is a higher-order Stokes multiplier which takes the values of $\mathcal{S}=1$ for $\text{Re}[Y]<0$ and $\mathcal{S}=0$ for $\text{Re}[Y]>0$. This change in $\mathcal{S}$ occurs smoothly across a boundary layer, surrounding the imaginary axis, of diminishing width as $n \to \infty$. The solution divergence \eqref{eq:LateSolIntro} is also associated with a divergent eigenvalue, of the form 
\begin{equation}\label{eq:LateEigIntro}
\lambda_{n} \sim
\left\{\begin{aligned}
\Big[ \delta_{0} \log{(n)}+\delta_{1} \Big]&\Gamma \Big(\frac{n+1}{2}+\alpha_0\Big)  \qquad \text{for $n$ even},\\
  \delta_{2}  &\Gamma \Big(\frac{n+1}{2}+\alpha_1 \Big) \qquad \text{for $n$ odd}.
\end{aligned}\right.
\end{equation}
Once these late-term components of the solution and eigenvalue are known, a procedure for the derivation of the exponentially-small components can be followed.

We now comment on the following pathologies related to \eqref{eq:LateSolIntro} and \eqref{eq:LateEigIntro}:
\begin{enumerate}
\item \emph{Divergent eigenvalues.} 

Although exponential asymptotics has been applied to other eigenvalue problems (cf. \cite{tanveer1987analytic}, \cite{kruskal_1991}, \cite{chapman2009exponential}, \cite{shelton2022exponential}), in such cases, the eigenvalue divergence has not been noted as significant. In the present work, the divergence of $\lambda_n$ affects the leading-order prediction of the eigenfunction divergence in \eqref{eq:LateSolIntro}, and is required to satisfy the associated boundary conditions on the late-term solution.

\item \emph{Spurious singularities in the late-term approximation}. 

It is known (cf. \cite{dingle_book}, \cite{berry_1989}, \cite{chapman_1998}) that typically, divergence of the late terms is captured by a factorial-over-power ansatz of the form displayed in \eqref{eq:typicalfactorial}. This factorial-over-power divergence is often taken as a universality of many problems in singularly perturbed asymptotics. 
However, we find that in the Hermite-with-pole problem, an additional singularity beyond that of $Y=-1$ is predicted by the divergent ansatz. This misleadingly suggests that the late-order divergence of the asymptotic series is attributed to a point where no singularity appears in the early orders. This unusual aspect is associated with the following item.

\item \emph{The higher-order Stokes phenomenon (HOSP).} 

The Hermite-with-pole problem exhibits a pathology where the anticipated Stokes phenomenon is suppressed in certain regions of the complex plane. This complexity is an example of the higher-order Stokes phenomena, for which a general analytic understanding from the viewpoint of the divergent series has remained elusive (c.f. \cite{howls_2004}, \cite{daalhuis2004higher}, \cite{body2005exponential}, \cite{chapman_2005}). Only the consequences of this phenomena will be discussed in this work, and we refer the reader to \cite{shelton2022HOSP} for a detailed derivation of HOSP from the perspective of the divergent series.

\item \emph{Atypical boundary layers in the late terms.} 

The na\"ive factorial-over-power divergence \eqref{eq:LateSolIntro} is unable to satisfy boundary condition \eqref{eq:bc2} at $Y=0$, due to the functional prefactor growing without bound as $Y \to 0$. A boundary layer of vanishing size as $n \to \infty$ must be introduced, in which the two divergences shown in \eqref{eq:LateSolIntro} interact, which also drives the HOSP of the previous point.

\item \emph{Even-and-odd pairing of the late terms.} 

Consecutive terms in the asymptotic expansion exhibit different singular behaviour at $Y=-1$: one is purely algebraic, and the other is the product of a logarithmic and an algebraic singularity. Consequently the late-term representation \eqref{eq:LateSolIntro} requires a different ansatz for $n$ even and $n$ odd. 
\end{enumerate}

It is the resolution of these complicated issues within that separates our work from the previous work by \cite{boyd1998sturm}. In the end, despite its misleadingly simple form, the Hermite-with-pole problem turns out to be quite a pathological investigation of beyond-all-orders asymptotics.

\section{An Initial Asymptotic Expansion} \label{sec:initial}
We begin by considering the asymptotic expansions
\begin{equation}
\label{eq:exp}
\psi(Y) = \sum _{n=0}^{\infty} \epsilon^n \psi_n(Y) \qquad \text{and} \qquad \lambda= \sum_{n=0}^{\infty} \epsilon^n \lambda_n.
\end{equation}
At leading order in equation \eqref{eq:hermite} we find the solution $\psi_0=C_0 Y^{-(1+\lambda_0)/2}$, where $C_0$ is a constant of integration. In general this solution is singular or contains a branch point at $Y=0$. In order to apply the leading-order boundary condition of $\psi_0(0)=1$ at the same location, a boundary layer should typically be considered. However, we can verify through an inner-matching procedure that the leading-order eigenvalue is $\lambda_0=-1$. Then the boundary condition at $Y=0$ gives $C_0=1$ and no boundary-layer theory is required. This yields our leading-order solution of 
\begin{equation} \label{eq:hermiteO1}
\psi_0= 1 \qquad \text{and} \qquad \lambda_0=-1.
\end{equation}
We emphasise that the singularity at $Y=0$ in the leading-order solution has been removed by the choice of the eigenvalue, $\lambda_0=-1$. A similar argument will be applied in subsequent orders to enforce regularity of the solution at $Y = 0$.

At the next order, $O(\epsilon)$, of equation \eqref{eq:hermite}, we find the solution
\begin{equation} \label{eq:hermiteO2}
\psi_1=C_1 + \frac{(1-\lambda_1)}{2} \log(Y) -\frac{1}{2} \log(1+Y),
\end{equation}
which contains singularities at both $Y=0$ and $Y=-1$. To apply the boundary condition $\psi_1(0)=0$, we require $\lambda_1=1$, which then determines the constant of integration as $C_1=0$. Thus, our $O(\ep)$ solution is
\begin{equation} \label{eq:hermiteO2sol}
\psi_1= -\frac{1}{2} \log(1+Y) \qquad \text{and} \qquad \lambda_1=1.
\end{equation}
Note that the above is singular at $Y=-1$. Since successive terms in the asymptotic series for $\psi$ in \eqref{eq:exp} rely on repeated differentiation of previous terms, the logarithmic singularity will result in the divergence of the series for $\psi_n$ as $n \to \infty$. It is this divergence that we wish to characterise. Note that in the $n \to \infty$ limit, on the assumption that $\psi_n$ is divergent, there exists a dominant balance between the two terms $\ep^2 \psi^{\prime \prime}$ and $-2 Y \psi^{\prime}$ of \eqref{eq:hermite}. Thus, we must continue to derive additional early orders of the solution until the effects of the $\ep^2 \psi^{\prime \prime}$ term become apparent. Since the singularity at $Y=-1$ in $\psi_1$ first appears at $O(\ep)$, the effects of this term will begin at $O(\ep^3)$.

The same procedure is applied at $O(\ep^2$) and $O(\ep^3$), for which we find the solutions 
\begin{subequations}
\begin{gather}\label{eq:hermiteO3sol}
\psi_2 = \frac{1}{8} \log^2(1+Y), \qquad \lambda_2=0,\\
\psi_3=-\frac{Y}{4(1+Y)}-\frac{1}{48}\log^3 (1+Y)-\frac{1}{4}\log (1+Y), \qquad \lambda_3=\frac{1}{2}.\label{eq:hermiteO4sol}
\end{gather}
\end{subequations}

Note that while the singularities at $Y=-1$ in $\psi_1$ and $\psi_2$ were logarithmic, the dominant singularity in $\psi_3$ is algebraic and of order unity. Typically the order of the singular behaviour of successive terms in the asymptotic series would increase linearly in a predictable fashion (see \emph{e.g.} the work by \cite{chapman_1998}). This is not the case for our current problem, which can be seen by progressing to the next order, which has the solution
\begin{equation} \label{eq:hermiteO5sol}
\psi_4=-\frac{\log(1+Y)}{8(1+Y)}-\frac{Y}{8(1+Y)}+\frac{\log^4 (1+Y)}{384}+\frac{\log^2 (1+Y)}{8} \quad \text{and} \quad \lambda_4=\frac{1}{4}.
\end{equation}

From \eqref{eq:hermiteO4sol} and \eqref{eq:hermiteO5sol}, we find the singular scalings, as $Y \to -1$, of 
\begin{equation} \label{eq:pairscaling}
\psi_3 \sim \frac{1}{4(1+Y)}  \qquad \text{and} \qquad
\psi_4 \sim \frac{-\log(1+Y)}{8(1+Y)}.
\end{equation}
From this, we anticipate that the singular behaviour as $Y \to -1$ of the asymptotic series will proceed in the pairwise fashion of
\begin{equation}
\label{eq:latepair}
\psi_{2k-1} =O \bigg( \frac{1}{(1+Y)^{k-1}} \bigg)\qquad \text{and} \qquad \psi_{2k} = O \bigg( \frac{\log(1+Y)}{ (1+Y)^{k-1}}\bigg)
\end{equation}
for integer $k \geq 2$, and hence the order of the algebraic
singularity increases every other term. As it turns out, the above
form in \eqref{eq:latepair}, which predicts the behaviour of the
late-order terms as $Y \to -1$ and $n \to \infty$ also hints at the
proper ansatz for $n \to \infty$ in general. In the late-term analysis
that follows we will employ separate divergent predictions for
$\psi_n$, distinguishing between the cases of $n$ even and $n$ odd.
The decoupling of the even and odd terms in the expansion as $n \rightarrow
\infty$ essentially
arises because (\ref{eq:hermite}) without the $\eps \psi/(1+Y)$ term
would have a  natural expansion in powers of $\eps^2$, but the
addition of this term forces an expansion in powers of $\eps$; similar
behaviour has been observed in \cite{chapman_1999}.

\section{Typical exponential asymptotics and the na\"ive divergence} \label{sec:expasym}
The goal of the exponential asymptotics procedure is to predict the exponentially-small eigenvalue and eigenfunction solutions. We shall see in \S\ref{sec:smoothing} that these exponentially-small terms are connected to the divergence of the expansion \eqref{eq:exp}. 

Our task in this section is to derive the analytical form of the late terms of \eqref{eq:exp} in the limit of $n \to \infty$. For this, we follow the procedure of introducing an ansatz for the factorial-over-power divergence. However, this ansatz, given in equation \eqref{eq:naiveansatz} below, takes an unusual form due to the inclusion of a $\log{(n)}$ divergent scaling for even values of $n$. It is demonstrated in \S\ref{sec:innersolouter}, through an inner analysis at the singularity, why the divergent ansatz must take this form.

At $O(\epsilon^n)$ in \eqref{eq:hermite}, we have
\begin{subequations}
\begin{equation}
\label{eq:late}
\psi_{n-2}^{\prime \prime} - 2Y \psi_{n}^{\prime} -\frac{Y}{1+Y}\psi_{n-1}= \lambda_3\psi_{n-3} + \cdots + \lambda_{n-1} \psi_1+ \lambda_n,
\end{equation}
and the boundary-condition of \eqref{eq:bc2} yields at $O(\ep^n)$
\begin{equation}
\label{eq:latebc}
\psi_{n}(0)=0.
\end{equation}
\end{subequations}
The late-order solutions, $\psi_n$, will contain a singularity at $Y=-1$ in the manner prescribed by equation \eqref{eq:latepair}. Moreover, since subsequent orders are determined by differentiation of earlier terms in the expansion, we anticipate that the divergence of the solution, introduced in \eqref{eq:LateSolIntro}, will be captured by the factorial-over-power ansatz,
\begin{equation} \label{eq:naiveansatz}
\psi_{n} \sim 
\left\{\begin{aligned}
\Big[ L(Y)\log{(n)} + Q(Y) \Big]\frac{\Gamma(\frac{n}{2} +\alpha_0)}{[\chi(Y)]^{n/2 +\alpha_0}}& \quad \text{for $n$ even},\\
R(Y)\frac{\Gamma(\frac{n}{2}+\alpha_1)}{[\chi(Y)]^{n/2+\alpha_1}}& \quad \text{for $n$ odd}.
\end{aligned}\right.
\end{equation}

As we have warned, the analysis to follow is quite involved. In essence, our first task is to derive the so-called \textit{na\"ive divergence} that appears in \eqref{eq:LateSolIntro} and above in \eqref{eq:naiveansatz}. This is performed in \S\ref{sec:homo} by neglecting the late-terms of the eigenvalue in the $O(\ep^n)$ equation. Before we do this, however, we shall motivate the unusual form of \eqref{eq:naiveansatz} in the next section by considering the outer limit of an inner solution at the boundary-layer near $Y=-1$. 

\subsection{Inner problem for the singularity of $Y=-1$} \label{sec:innernew}
First, we note that the early orders of expansion \eqref{eq:exp} reorder as we approach the singularity at $Y=-1$.
Instead of consecutive terms in the outer expansion reordering, those with an odd and even powers of $\epsilon$ will reorder amongst themselves. For instance, the reordering occurs between odd terms for $\ep^3 \psi_3 \sim \ep^5 \psi_5$ and even terms for $\ep^4\psi_4 \sim \ep^6 \psi_6$. Since $\psi_3 \sim  (1+Y)^{-1}$ and $\psi_5 \sim (1+Y)^{-2}$ from the singular behaviour introduced in equation \eqref{eq:latepair}, we balance $(1+Y)^{-1} \sim \ep^2 (1+Y)^{-2}$ to find the width of the boundary layer to be of $O(\ep^2)$. The same width is found by considering the even reordering. We thus introduce the inner-variable, $\hat{y}$, by setting
\begin{equation}
\label{eq:innery}
1+Y=\ep^2 \hat{y},
\end{equation}
with $\hat{y}$ of $O(1)$ in the inner region.
The inner equation may then be derived by substituting for $\hat{y}$, giving
\begin{equation}
\label{eq:innerEq}
\dd{^2 \hat{\psi}}{\hat{y}^2} +2(1-\ep^2 \hat{y}) \dd{\hat{\psi}}{\hat{y}} + \frac{\ep \hat{\psi}}{\hat{y}} = \ep^2 (1+\lambda) \hat{\psi},
\end{equation}
where we denote the inner solution by $\hat{\psi}$.

\subsubsection{Inner limit of the early orders}\label{sec:innerearlylim}
To motivate the correct form for the inner solution, we take the inner limit of the outer solution by substituting for $\hat{y}$ and expanding as $\ep \to 0$. This yields
\begin{equation} \label{eq:innerLim}
\begin{aligned}
\psi_{\text{outer}}\sim 1 - \ep \log{(\ep)} + \ep \bigg[-\frac{\log{(\hat{y})}}{2}+\frac{1}{4\hat{y}}+\cdots\bigg] + \frac{\ep^2 \log^2{(\ep)}}{2} \qquad \qquad \qquad \quad \\
+ \ep^2 \log{(\ep)} \bigg[\frac{\log{(\hat{y})}}{2} - \frac{1}{4 \hat{y}} +\cdots\bigg]
+\ep^2 \bigg[\frac{\log^2{(\hat{y})}}{8} -\frac{\log{(\hat{y})}}{8\hat{y}}+\frac{1}{8\hat{y}} + \cdots \bigg] +\cdots.
\end{aligned}
\end{equation}

\subsubsection{Outer limit of the inner solution}\label{sec:innersolouter}
In Appendix \ref{sec:innersolnew}, we solve the inner equation
\eqref{eq:innerEq} by considering an inner solution, motivated by
\eqref{eq:innerLim}, of the form $\hat{\psi}= \hat{\psi}_0
+\ep\log{(\ep)} \hat{\psi}_{(1,1)}+ \ep \hat{\psi}_1 +\ep^2
\log^2{(\ep)} \hat{\psi}_{(2,2)}+\ep^2 \log{(\ep)} \hat{\psi}_{(2,1)}+
\ep^2 \hat{\psi}_2+\cdots$.
We write the inner solution from \eqref{eq:innerSol} in outer
variables by substituting for $\hat{y}=(1+Y)/{\ep^2}$ to give 
the outer limit (of the first six terms of the inner series) as
\begin{equation}\label{eq:innerouterlim}
\begin{aligned}
\hat{\psi} \sim  1 - \ep \frac{\log{(1+Y)}}{2} + \ep^2 \frac{\log^2{(1+Y)}}{8} +  \sum_{k=1}^{\infty} \frac{\ep^{1+2k}}{2}\frac{\Gamma(k)}{[2(1+Y)]^k}&\\
+\sum_{k=1}^{\infty}\frac{\ep^{2+2k}}{4}  \frac{[4b_k-\log{(1+Y)}\Gamma(k)]}{[2(1+Y)]^{k}}&.
\end{aligned}
\end{equation}
The divergent constant $b_k$ is determined by the recurrence relation
\eqref{eq:akeq}, which may be solved in the limit of $k \to \infty$ (as performed in \eqref{eq:bksollim}) to give $b_k \sim \tfrac{1}{2}(\log{(k)} + \gamma)\Gamma(k)$. It is this extra factor of $\log{(k)}$ in the expansion for $b_k$ that causes the unusual $\log{(n)}$ divergent form introduced in \eqref{eq:naiveansatz}.
In order to compare \eqref{eq:innerouterlim} with the late-terms of
the outer solution at $O(\ep^n)$, we substitute $n=1+2k$ for the first
sum on the right-hand side of \eqref{eq:innerouterlim} and $n=2+2k$
for the second. The $O(\ep^n)$ of this outer limit is then
\begin{equation} \label{eq:innerouterlimn}
\hat{\psi} \sim 
\left\{\begin{aligned}
\ep^{n}  \bigg[\frac{1}{2}\log{(n)}+\frac{\gamma-\log{(2)}}{2}-\frac{1}{4}\log{(1+Y)}\bigg]\frac{\Gamma(\frac{n}{2}-1)}{[2(1+Y)]^{\frac{n}{2}-1}}& \quad \text{for $n$ even},\\
 \frac{\ep^{n}}{2} \frac{\Gamma(\frac{n-1}{2})}{[2(1+Y)]^{\frac{n-1}{2}}}& \quad \text{for $n$ odd},
\end{aligned}\right.
\end{equation}
where we expanded $b_{n/2-1} \sim \tfrac{1}{2}[\log{(n)}+\gamma-\log{(2)}]\Gamma(\frac{n}{2}-1)$ for $n \to \infty$ as in \eqref{eq:bksollim}. 
Equation \eqref{eq:innerouterlimn} motivates the slightly unusual form of the  factorial-over-power ansatz we had previously introduced in \eqref{eq:naiveansatz}. We are now ready to return to study the divergence of the outer solution. 

\subsection{Divergence of the homogeneous late-term equation} \label{sec:homo}
In this section, we derive the {\it na\"ive divergence}, which is obtained as a solution to the the $O(\ep^n)$ equation \eqref{eq:late} when the late-terms of the eigenvalue are neglected. We thus study the equation
\begin{equation}
\label{eq:latehom}
\psi_{n-2}^{\prime \prime} - 2Y \psi_{n}^{\prime} -\frac{Y}{1+Y}\psi_{n-1}= \lambda_3\psi_{n-3} + \cdots,
\end{equation}
where the lower order terms on the right hand side are of orders $\psi_{n-4}$, $\psi_{n-5}$, and so forth.
Later in \S\ref{sec:constchisol}, we demonstrate that the late-terms
of the eigenvalue produces particular solutions that are subdominant
as $n \to \infty$ near the singularity of $Y=-1$, but which
  are crucially responsible for the higher-order Stokes phenomenon.

Substituting the factorial-over-power ansatz \eqref{eq:naiveansatz}
into the homogeneous equation \eqref{eq:latehom}, the dominant terms
in the equation are  of
$O(\log{(n)}\Gamma({n}/{2}+\alpha_0+1)/\chi^{{n}/{2}+\alpha_0+1})$ for
$n$ even and $O(\Gamma(n/2+\alpha_1+1)/\chi^{n/2+\alpha_1+1})$ for $n$
odd.
Dividing out this dominant behaviour gives terms of order $O(1)$, $O(n^{-1})$, etc. for $n$ odd, and $O(1)$, $O(\log^{-1}{n})$, $O(n^{-1})$, $O(n^{-1}\log^{-1}{n})$ etc. for $n$ even. At leading order as $n \to \infty$,  both cases give
\begin{equation}
\label{eq:chieq}
\chi^{\prime} (\chi^{\prime}+2Y)=0.
\end{equation}
The singular behaviour of $\psi_n$ will be captured by the non-trivial
solution, $\chi^{\prime}=-2Y$. Since we require $\chi(-1)=0$ in order to
match with the inner solution near the singularity from
\eqref{eq:innerouterlimn}, we find
\begin{equation} \label{eq:mychi}
\chi(Y)=1-Y^2.
\end{equation}
Equations for the prefactor functions $L$, $Q$, and $R$ are found at the following orders of $n$ in equation \eqref{eq:latehom}. Since even and odd components of the divergence now interact, between $\psi_n$ and $\psi_{n-1}$ for instance, it is necessary to specify
\begin{equation}
\label{eq:gamma1}
\alpha_1 - \alpha_0 = 1/2,
\end{equation}
in keeping with the different rates of divergence in (\ref{eq:innerouterlimn}).
At $O(n^{-1})$ for $n$ even we find an equation for $L(Y)$. Similarly, the $R(Y)$ and $Q(Y)$ equations are found at $O(n^{-1}\log^{-1}{n})$ for the cases of $n$ odd and $n$ even, respectively. These equations are
\begin{subequations}
\begin{align}\label{eq:prefactoreq}
&L^{\prime}(Y)+\frac{1}{Y}L(Y)=0, \qquad  R^{\prime}(Y)+\frac{1}{Y}R(Y)=0,\\
\label{eq:prefactoreqQ}
&Q^{\prime}(Y)+\frac{1}{Y}Q(Y)=\frac{R(Y)}{2(1+Y)} + \frac{2Y L(Y)}{1-Y^2},
\end{align}
\end{subequations}
which may be integrated directly to find the solutions
\begin{subequations}
\begin{align}\label{eq:prefactor}
  L(Y)=&\frac{\Lambda_{\text{L}}}{Y},     \qquad  R(Y)=\frac{\Lambda_{\text{R}}}{Y},\\
\label{eq:prefactorQ}
Q(Y)=&\frac{\Lambda_{\text{Q}}}{Y}+\frac{\Lambda_{\text{R}}}{2Y} \log(1+Y)-\frac{\Lambda_{\text{L}}}{Y}\log{(1-Y^2)},
\end{align}
\end{subequations}
where  $\Lambda_{\text{L}}$, $\Lambda_{\text{R}}$, and $\Lambda_{\text{Q}}$ are constants of integration.

Substitution of solutions \eqref{eq:prefactor} and \eqref{eq:prefactorQ} into the ansatz \eqref{eq:naiveansatz} gives our divergent prediction for $\psi_n$, with $n \to \infty$ as
\begin{equation} \label{eq:naive}
\psi_{n} \sim 
\left\{\begin{aligned}
\bigg[ \frac{\Lambda_{\text{L}}}{Y}\log{(n)} + \bigg(\frac{\Lambda_{\text{Q}}}{Y}+\frac{\Lambda_{\text{R}}}{2Y}\log(1+Y) \qquad \qquad \qquad & \\ - \, \frac{\Lambda_{\text{L}}}{Y}\log{(1-Y^2)}\bigg)\bigg]\frac{\Gamma(\frac{n}{2} +\alpha_0)}{(1-Y^2)^{n/2 +\alpha_0}}& \qquad  \text{for $n$ even},\\
\frac{\Lambda_{\text{R}}}{Y}
\frac{\Gamma(\frac{n}{2}+{\alpha_0+\frac{1}{2}})}{(1-Y^2)^
  {n/2+{\alpha_0+1/2}}}& \qquad \text{for $n$ odd}.
\end{aligned}\right.
\end{equation}
We refer the late-order form of \eqref{eq:naive} as corresponding to the \textit{na\"ive divergence}, for which two noticeable issues are present: 
\begin{enumerate}
\item The boundary condition, $\psi_n(0)=0$, is unable to be satisfied as our current form is unbounded at $Y=0$;
\item There are additional locations at which the singulant, $\chi(Y)$, is equal to zero. Since $\chi(Y)=1-Y^2$, our late term expression predicts singularities at both $Y=-1$ and $Y=1$. This is in contrast to the early orders of the expansion, which are singular at $Y=-1$ only.
\end{enumerate}

The first of these issues will be resolved in \S\ref{sec:reorder}. There, we demonstrate that as $n \to \infty$, a boundary layer emerges in the late-order solution near $Y = 0$. This boundary layer is of diminishing width as $n \to \infty$. A matched asymptotic approach then allows us to develop an inner solution that satisfies the boundary condition of $\psi_n(0) = 0$.
Regarding the the second issue, the late terms \eqref{eq:naive} in fact switch off across a higher-order Stokes line along the imaginary axis. This is known as the higher-order Stokes phenomenon, which is in fact generated by the singularity discussed in item 1. For a derivation of this phenomenon from the perspective of the divergent series, we refer the reader to the work by \cite{shelton2022HOSP}.

\subsection{Determination of the unknown constants}
It remains to find values for the constants $\Lambda_{\text{L}}$,
$\Lambda_{\text{R}}$, $\Lambda_{\text{Q}}$ and $\alpha_0$, that appear
in the  late-term solution for $\psi_n$ in \eqref{eq:naive}. These are determined through matching with the outer limit of the inner solution about the singularity at $Y=-1$ given in equation \eqref{eq:innerouterlimn}. Expanding the outer solution for $\psi_n$ from \eqref{eq:naive} as $Y \to -1$, we have
\begin{equation} \label{eq:naivetosing}
\psi_{n} \sim 
\left\{\begin{aligned}
\bigg[ -\Lambda_{\text{L}} \log{(n)} + \bigg(\Lambda_{\text{L}}\log{(2)}-\Lambda_{\text{Q}}&\\
+\Big[ \Lambda_{\text{L}}-\frac{\Lambda_{\text{R}}}{2}\Big]\log(1+Y) \bigg)&\bigg]\frac{\Gamma(\frac{n}{2} +\alpha_0)}{[2(1+Y)]^{n/2 +\alpha_0}} \quad \text{for $n$ even},\\
-\Lambda_{\text{R}} &\frac{\Gamma(\frac{n}{2}+{\alpha_0+\frac{1}{2}})}{[2(1+Y)]^{n/2+{\alpha_0+1/2}}} \quad \text{for $n$ odd}.
\end{aligned}\right.
\end{equation}
This form may now be compared to the outer limit of the inner solution in \eqref{eq:innerouterlimn} to find
\begin{equation}\label{eq:consts}
\Lambda_{\text{R}} = -\frac{1}{2}, \qquad \Lambda_{\text{L}}=-\frac{1}{2}, \qquad \Lambda_{\text{Q}}=-\frac{\gamma}{2}, \qquad \alpha_0=-1,
\end{equation}
where $\gamma \approx 0.577$ is the Euler-Macheroni constant.

\section{Late-term divergence of the eigenvalue expansion}
\subsection{The boundary layer near $Y=0$} \label{sec:jonYzero}
We saw in \S\ref{sec:initial} that each term in the expansion of the
eigenvalue was determined by imposing that the outer solution had no
singularity at $Y=0$. We can find this expansion more readily by
considering a local expansion in the vicinity of $Y=0$.
Writing $Y  = \eps y$ we find
\[ \sdd{\psi}{y} - 2 y \dd{\psi}{y}  + \frac{\eps\psi}{1+\eps y} = (\lambda+1)
  \psi.\]
Expanding in powers of $\eps$ as usual,
\[ \psi = \sum_{n=0}^\infty \eps^n \psi_n\]
gives
\begin{equation}
  \sdd{\psi_n}{y} - 2 y \dd{\psi_n}{y}  = - \sum_{k=1}^n (-y)^{k-1}
  \psi_{n-k} + \sum_{k=1}^{n}\la_k \psi_{n-k},\label{ynear0}
  \end{equation}
where we have expanded
\[  \frac{\eps}{1+\eps y}  
  =\sum_{n=1}^\infty \eps^n (-y)^{n-1}  .
\]
We find $\psi_0=1$, $\psi_1=0$, $\la_1=1$, and in general
\[\psi_n = \sum_{m=1}^{n-1} a_{m,n} y^m \qquad \mbox{ for }n \geq 2,\]
where the series coefficient satisfies the recurrence relation
\begin{equation}
     2 r a_{r,n}  = (r+2)(r+1)a_{r+2,n}- \sum_{k=1}^{n-r} \la_ka_{r,n-k} 
   + \sum_{k=1}^{r+1} (-1)^{k-1} a_{r+1-k,n-k} \label{recurrence}
 \end{equation}
 with  $a_{r,n}=0$ if $r \geq n-1$ or $r=0$.
 It is straightforward to solve (\ref{recurrence}) numerically,
 stepping down from $r=n-1$ to $r=0$ for each $n$. When $r=0$ the
 left-hand side is zero; that the right-hand side must vanish then
 gives the equation for $\la_n$, which is
 \[ \la_n = 2 a_{2,n}.\]
 These numerical solutions are later compared to the divergent prediction for $\lambda_n$ in figure \ref{fig:EigScaling}. 
 
 It is not so straightforward to determine the divergence of $\la_n$
 as $n \rightarrow \infty$ from (\ref{recurrence}), but we can make
 some progress by observing that the solution of the homogeneous
 adjoint to (\ref{ynear0}) is $\e^{-y^2}$. Multiplying by this and
 integrating gives
\begin{eqnarray*}
\la_n \sqrt{\pi} &=&   \sum_{k=1}^n \int_{-\infty}^\infty
\e^{-y^2}(-y)^{k-1}\psi_{n-k}\,\d y 
- \sum_{k=1}^{n-1}\la_k \int_{-\infty}^\infty
\e^{-y^2}\psi_{n-k}\, \d y\\
& \sim & \frac{(1 - (-1)^n)}{2} \Gamma(n/2) + \cdots.
\end{eqnarray*}
When $n$ is odd this gives
\begin{equation}
  \la_n = \frac{1}{\sqrt{\pi}} \Gamma(n/2).\label{lanodd}
  \end{equation}
However, when $n$ is even the first term vanishes, and the correction
term is much harder to determine.

\subsection{Solution divergence forced by the eigenvalue} \label{sec:constchi}
We now consider the particular solution of (\ref{eq:late}) generated by the
divergent eigenvalue expansion $\la_n$. We will see (and motivated by
(\ref{lanodd})) that the correct form of the eigenvalue divergence is 
\begin{equation}\label{eq:LateEig}
\lambda_{n} \sim
\left\{\begin{aligned}
\Big[ \delta_{0} \log{(n)}+\delta_{1}\Big]\Gamma \Big(\frac{n-1}{2}\Big)&  \quad \text{for $n$ even},\\
  \delta_{2}  \Gamma \Big(\frac{n}{2} \Big)&  \quad \text{for $n$ odd},
\end{aligned}\right.
\end{equation}
where we expect to find $\delta_2=1/\sqrt{\pi}$.
We find that this generates a particular solution in $\psi_n$ of the
form
\begin{equation}\label{eq:ConstChiAn}
\psi_{n}(Y)  \sim
\left\{\begin{aligned}
 \bigg[Q^{(\lambda_n)}_{0}(Y)\log^2{(n)}+ Q^{(\lambda_n)}_{1}(Y) \log(n)+Q^{(\lambda_n)}_{2}(Y) \bigg] &\Gamma \Big(\frac{n-1}{2} \Big)  \quad \text{for $n$ even,} \\
 \bigg[ R^{(\lambda_n)}_{1}(Y) \log(n)+R^{(\lambda_n)}_{2}(Y) \bigg] &\Gamma \Big(\frac{n}{2} \Big) \quad \qquad \text{for $n$ odd.} 
\end{aligned}\right.
\end{equation}

Substituting ansatz \eqref{eq:ConstChiAn} into the $O(\ep^n)$ equation \eqref{eq:late}, we divide out by the dominant behaviour, which is $\log^2{(n)}\Gamma((n-1)/2)$ for $n$ even and $\log{(n)}\Gamma(n/2)$ for $n$ odd. At $O(n^{0})$ for $n$ odd and $n$ even, we then find 
\begin{equation}
\label{eq:OuterConstEq1}
R_{1}^{(\lambda_n)\prime}(Y)=0,  \qquad Q_{0}^{(\lambda_n)\prime}(Y)=0,
\end{equation}
with solution
\begin{equation}
\label{eq:OuterConstSol1}
R^{(\lambda_n)}_{1}(Y)=A_1 , \qquad Q^{(\lambda_n)}_{0}(Y)=B_0,
\end{equation}
where $A_1$ and $B_0$ are constants. Next, at $O(\log^{-1}{(n)})$, for
$n$ odd and even respectively, we find
\begin{equation}
\label{eq:OuterConstEq2}
R_{2}^{(\lambda_n)\prime}(Y)=-\frac{\delta_{2}}{2Y}  \qquad \text{and} \qquad Q_{1}^{(\lambda_n)\prime}(Y)=-\frac{\delta_{0}}{2Y}-\frac{A_1}{2(1+Y)},
\end{equation}
with solution
\begin{equation}
\label{eq:OuterConstSol2}
R^{(\lambda_n)}_{2}(Y)=A_2-\frac{\delta_{2}}{2}\log{(Y)},  \quad  Q^{(\lambda_n)}_{1}(Y)=B_1-\frac{\delta_{0}}{2}\log{(Y)}-\frac{A_1}{2}\log{(1+Y)},
\end{equation}
where $A_2$ and $B_1$ are constants. At the next order of $O(\log^{-2}(n))$ for $n$ even, we find
\begin{equation}
\label{eq:OuterConstEq3}
Q_2^{(\lambda_n)\prime}(Y)=-\frac{\delta_{1}}{2Y}- \frac{\delta_{2}}{2Y}\psi_1(Y)-\frac{R^{(\lambda_n)}_2(Y)}{2(1+Y)},
\end{equation}
with solution
\begin{equation}\label{eq:ConstSol3}
Q^{(\lambda_n)}_2(Y)=B_2 - \frac{\delta_{1}}{2} \log(Y) - \frac{A_2}{2}\log(1+Y)+\frac{\delta_{2}}{4}\log(Y)\log(1+Y).
\end{equation}
Overall, the divergence of
$\psi_n$ is given by combining \eqref{eq:naive} with
\eqref{eq:ConstChiAn} to give
\begin{equation}\label{eq:FullOuter}
\psi_{n} \sim 
\left\{\begin{aligned}
\Big[ L(Y) \log{(n)} + Q(Y) \Big]\frac{\Gamma(\frac{n}{2} -1)}{\chi^{n/2 -1}} \qquad \qquad ~~\qquad \qquad \quad&\\
+\Big[ Q^{(\lambda_n)}_0(Y)\log^2{(n)}+Q^{(\lambda_n)}_1(Y) \log(n)+Q^{(\lambda_n)}_2(Y) \Big] \Gamma \Big(\frac{n-1}{2}\Big)& \quad \text{for $n$ even},\\
R(Y) \frac{\Gamma(\frac{n-1}{2})}{\chi^{(n-1)/2}}+\Big[ R^{(\lambda_n)}_1(Y) \log(n)+R^{(\lambda_n)}_2(Y)  \Big] \Gamma \Big(\frac{n}{2} \Big)& \quad  \text{for $n$ odd}.
\end{aligned}\right.
\end{equation}
In the next section, we demonstrate how  these divergences
interact in a boundary layer near $Y=0$, justifying the ansatzes
\eqref{eq:LateEig} and \eqref{eq:ConstChiAn}, resolving issues 1 and
2 in \S\ref{sec:homo}, and  determining the
coefficients $\delta_0$, $\delta_1$ and $\delta_2$.

\section{The late-term boundary layer at $Y=0$}\label{sec:constchisol}
Recall that in the early orders of the expansion, each order of the
eigenvalue was determined by enforcing the boundary condition at
$Y=0$. However, late term expansion  \eqref{eq:FullOuter}
is unbounded at $Y=0$ and cannot satisfy the condition  $\psi_n(0)=0$. 
If we continue the expansion \eqref{eq:FullOuter} to higher orders (in
$1/n$)  we find that the
singularity at leading order (for instance $R_0 \sim Y^{-1}$) forces a
stronger singularity at the next order (so that $R_1 \sim
Y^{-3}$). Thus, this series reorders as $Y \to 0$, so that there is a
boundary layer in the late-term approximation near $Y=0$, for which an inner
analysis is required.
Note the distinction between this boundary layer and that of \S
\ref{sec:jonYzero}. There the boundary layer was due a nonuniformity
in the expansion of $\psi$ in $\eps$, and involved rescaling $Y$ with
$\eps$. Here the boundary later is due to a nonuniformity in the
expansion of $\psi_n$ in $n$, and involves rescaling $Y$ with $n$.

\subsection{Reordering of the late-terms as $Y \to 0$} \label{sec:reorder}
In order to determine the width of this boundary layer in the late-term solution, we introduce in Appendix \ref{sec:LowerOrderNaive} a factorial-over-power ansatz of the form
\begin{equation} \label{eq:16ab}
\psi_{n} \sim
\left\{\begin{aligned}
 \bigg[L_0(Y)\log{(n)}+Q_0(Y) + \frac{\log{(n)}}{n}L_1(Y)+&\cdots \bigg] \frac{\Gamma(\frac{n}{2} -1)}{\chi^{n/2-1}} \quad \text{for $n$ even}, \\
 \bigg[R_0(Y) +\frac{\log{(n)}}{n}M_1(Y)+ \frac{R_1(Y)}{n} +&\cdots\bigg]\frac{\Gamma(\frac{n-1}{2} )}{\chi^{(n-1)/2}} \quad \text{for $n$ odd}.
\end{aligned}\right.
\end{equation}
Here, the leading order solutions of $L_0(Y)$, $R_0(Y)$, and $Q_0(Y)$ are the same as $L(Y)$, $R(Y)$, and $Q(Y)$ derived previously in \eqref{eq:prefactor} and \eqref{eq:prefactorQ}. The solutions of $M_1(Y)$, $L_1(Y)$, and $R_1(Y)$ are given in equations \eqref{eq:M1sol} and \eqref{eq:R1sol}. For the purposes of observing the reordering of these series near $Y=0$, it is sufficient to display only their singular behaviour here, which is given by
\begin{equation}\label{eq:19}
L_0 \sim \frac{\Lambda_{\text{L}}}{Y}, \qquad  L_1 \sim  \frac{\Lambda_{\text{L}}}{Y^3}, \qquad
R_0 \sim \frac{\Lambda_{\text{R}}}{Y}, \qquad R_1 \sim \frac{\Lambda_{\text{R}}}{Y^3}.
\end{equation}
The series expansions of $\psi_n$ reorder when the two consecutive
terms in each of \eqref{eq:16ab} are of the same order as $n \to
\infty$. Since this occurs for
$Y = O(n^{-1/2}),$
we introduce the inner variable
$\bar{y}=n^{1/2} Y$.
Substituting this inner variable into the $O(\epsilon^n)$ equation
\eqref{eq:late} gives the inner equation as
\begin{equation}
\label{eq:InnerEq}
n \dd{^2\bar{\psi}_{n-2}}{\bar{y}^2}-2\bar{y} \dd{\bar{\psi}_n}{\bar{y}} + \frac{\bar{y}}{n^{1/2}}\bigg(1+\frac{\bar{y}}{n^{1/2}}\bigg)^{-1} \bar{\psi}_{n-1}=\lambda_3 \bar{\psi}_{n-3}+ \cdots +\lambda_{n-1}\bar{\psi}_{1} +\lambda_n,
\end{equation}
where $\bar{\psi}_{1}=-\frac{1}{2} \log(1+n^{-1/2} \bar{y}) \sim - \frac{1}{2}n^{-1/2}\bar{y}$.

\subsection{Inner limit of the late-term divergence}
We now take the inner limit of the outer divergent solution to
motivate the correct form for the inner solution. We begin by
substituting  the inner variable $\bar{y}$ in the
na\"ive divergence \eqref{eq:naive} and taking the limit  $n \to \infty$.  For
the singulant  we find
\begin{equation} \label{eq:InnerLimChi}
(1-Y^2)^{-n/2} = \bigg(1-\frac{\bar{y}^2}{n}\bigg)^{-n/2} \sim ~ \mathrm{e}^{\bar{y}^2/2} \quad \text{as} \quad n \to \infty.
\end{equation}
Furthermore, the scaling of $Q(Y) \sim Y^{-1}$ will increase the
argument of the gamma function by one half. Together we find
\begin{equation} \label{eq:InnerLimNaive}
\psi_n \, \sim
\left\{\begin{aligned}
\bigg[-\frac{\log{(n)}}{\sqrt{2}}-\frac{\gamma}{\sqrt{2}}\bigg]\frac{\mathrm{e}^{\bar{y}^2/2}}{\bar{y}} \Gamma \Big(\frac{n-1}{2}\Big) 
&\quad \text{for $n$ even},\\
-\frac{1}{\sqrt{2}}\frac{\mathrm{e}^{\bar{y}^2/2}}{\bar{y}} \Gamma \Big(\frac{n}{2} \Big) 
&\quad \text{for $n$ odd}.
\end{aligned}\right.
\end{equation}

We now take the inner limit of the particular solution  generated by
the divergent eigenvalue \eqref{eq:ConstChiAn}, which yields
\begin{equation}\label{eq:ConstInnLim}
\psi_{n}  \sim
\left\{\begin{aligned}
 \bigg[\bigg(B_0 + \frac{\delta_{0}}{4} \bigg) \log^2{(n)}+ \bigg(B_1+\frac{\delta_{1}}{4} -\frac{\delta_{0}}{2}\log{(\bar{y})}\bigg)\log(n) \quad ~ & \\
 +\bigg( B_2 - \frac{\delta_{1}}{2} \log(\bar{y}) \bigg) \bigg]\Gamma \Big(\frac{n-1}{2} \Big) \quad &\text{for $n$ even}, \\
 \bigg[ \bigg(A_1+\frac{\delta_{2}}{4} \bigg)\log(n)+ \bigg( A_2 - \frac{\delta_{2}}{2} \log(\bar{y}) \bigg)\bigg] \Gamma \Big(\frac{n}{2} \Big) \quad &\text{for $n$ odd}.
\end{aligned}\right.
\end{equation}
Together we find for $n$ even
\begin{subequations}\label{eq:Innertotalboth}
\begin{equation}\label{eq:InnerLimEven}
\begin{aligned}
\psi_{n} \sim
\bigg[\bigg( B_0 + \frac{\delta_{0}}{4} \bigg)\log^2{(n)}+
 \bigg(-\frac{\e^{\bar{y}^2/2}}{\sqrt{2}\bar{y}}+B_1+\frac{\delta_{1}}{4}-\frac{\delta_{0}}{2}\log{(\bar{y})}\bigg)\log(n) \\
 +\Biggl( \frac{ -\gamma \e^{\bar{y}^2/2}}{\sqrt{2}\bar{y}} + B_2 -\frac{\delta_{1}}{2}\log{(\bar{y})} \bigg)\bigg] \Gamma \bigg( \frac{n-1}{2}\bigg),
 \end{aligned}
\end{equation}
and for $n$ odd
\begin{equation}\label{eq:InnerLimOdd}
\psi_{n} \sim
 \bigg[\bigg(A_1+\frac{\delta_{2}}{4} \bigg)\log(n) + 
 \Biggl( -\frac{\e^{\bar{y}^2/2}}{\sqrt{2}\bar{y}} + A_2 -\frac{\delta_{2}}{2}\log{(\bar{y})} \bigg)\bigg] \Gamma\bigg(\frac{n}{2}\bigg).
\end{equation}
\end{subequations}

\subsection{An inner solution} \label{sec:InnerSol}
We now look for a solution to the inner equation
\eqref{eq:InnerEq}. Motivated  by the form of the inner limit in \eqref{eq:Innertotalboth}, we make the ansatz
\begin{equation}\label{eq:InnerAn}
\bar{\psi}_{n} \sim
\left\{\begin{aligned}
\Big[ \bar{L}(\bar{y}) \log{(n)}+\bar{Q}(\bar{y}) \Big]\Gamma \Big(\frac{n-1}{2}\Big)&  \quad \text{for $n$ even},\\
  \bar{R}(\bar{y})  \Gamma \Big(\frac{n}{2} \Big)&  \quad  \text{for $n$ odd}.
\end{aligned}\right.
\end{equation}
Substituting \eqref{eq:InnerAn}  and \eqref{eq:LateEig} into  \eqref{eq:InnerEq}, and isolating the dominant factorial divergence of $\Gamma (\frac{n}{2})$ for $n$ odd and $\Gamma (\frac{n-1}{2}) $ for $n$ even, yields at leading order the equations
\begin{equation}
  \label{eq:IN3b}
  \bar{R}^{\prime \prime} - \bar{y} \bar{R}^{\prime} =
  \frac{\delta_{2}}{2},  \qquad
  \bar{L}^{\prime \prime} - \bar{y} \bar{L}^{\prime} = \frac{\delta_{0}}{2},  \qquad
\bar{Q}^{\prime \prime} - \bar{y} \bar{Q}^{\prime} =
\frac{\delta_{1}}{2}. 
\end{equation}
These three equations all have solutions of a similar form. We will
now focus on the  equation for $\bar{R}$, and adapt the following
results analogously  for $\bar{L}$ and $\bar{Q}$. Integrating  (\ref{eq:IN3b}) we find  
\begin{equation} \label{eq:IN4Q1}
\bar{R}(\bar{y}) = \bar{B}_{{R}}+\bar{A}_{{R}} \int_{0}^{\bar{y}} \e^{t^2/2} \, \de{t} + \frac{\delta_{2}}{2}\int_{0}^{\bar{y}}\e^{t^2/2}   \bigg[ \int_{0}^{t} \e^{-p^2/2} \de{p} \bigg] \de{t}, 
\end{equation}
with constants of integration $\bar{A}_{{R}}$ and
$\bar{B}_{{R}}$. We are now able to apply the condition
$\bar{\psi}_n(0) = 0$ (resolving issue 1 of \S\ref{sec:homo}), which gives
$\bar{B}_{{R}}=0$.
The remaining constants are determined by matching with the the outer solution.

We see that in the outer limit of $|\bar{y}| \to \infty$ (\ref{eq:IN3b}a)
itself exhibits Stokes phenomenon.
There is a Stokes line on the imaginary axis, across which
the asymptotic behaviour of the term proportional to $\delta_2$
changes from
\[ \frac{\log(-\bar{y})}{2} +\cdots -
  \frac{\pi^{1/2}}{2^{3/2} }\frac{\e^{\bar{y}^2/2}}{\bar{y}}\qquad
  \mbox{ to }
  \qquad 
\frac{\log(-\bar{y})}{2} +\cdots +
\frac{\pi^{1/2}}{2^{3/2}}\frac{\e^{\bar{y}^2/2}}{\bar{y}}.
\]
This is an example of what is known as higher-order Stokes phenomenon,
which is a Stokes phenomenon in the asymptotic approximation of the
late terms of the expansion.
Additionally there is a second Stokes line on the real axis, across which the
asymptotic behaviour of the term proportional to
$\bar{A}_{{R}}$ picks up an additional constant (the
complementary function is just an error function of imaginary
argument).
Altogether, on the real axis, as  $\bar{y} \to \infty$,
\begin{equation}\label{eq:InnerInf}
\bar{R}(\bar{y})  \sim
\bigg[\bar{A}_{{R}}+\frac{1}{2}\Big(\frac{\pi}{2}\Big)^{\tfrac{1}{2}}\delta_{2}\bigg]\frac{\e^{\bar{y}^2/2}}{\bar{y}}
+\cdots -\frac{\delta_{2}}{4}\bigg( \log(2)+\gamma+\log(-\bar{y}^2)+\cdots \bigg),
\end{equation}
where $\gamma \approx 0.577$ is the Euler-Mascheroni constant, while
as  $\bar{y} \to -\infty$,
\begin{equation}\label{eq:InnerInf2}
\bar{R}(\bar{y})  \sim
\bigg[\bar{A}_{{R}}-\frac{1}{2}\Big(\frac{\pi}{2}\Big)^{\frac{1}{2}}\delta_{2}\bigg]\frac{\e^{\bar{y}^2/2}}{\bar{y}}
+\cdots -\frac{\delta_{2}}{4}\bigg( \log(2)+\gamma+\log(-\bar{y}^2) +\cdots\bigg).
\end{equation}
Exactly similar expressions hold for $\bar{L}$ and $\bar{Q}$.

\subsection{Matching} \label{sec:matching}
\begin{figure}[htpb]
	\centering
	\includegraphics[]{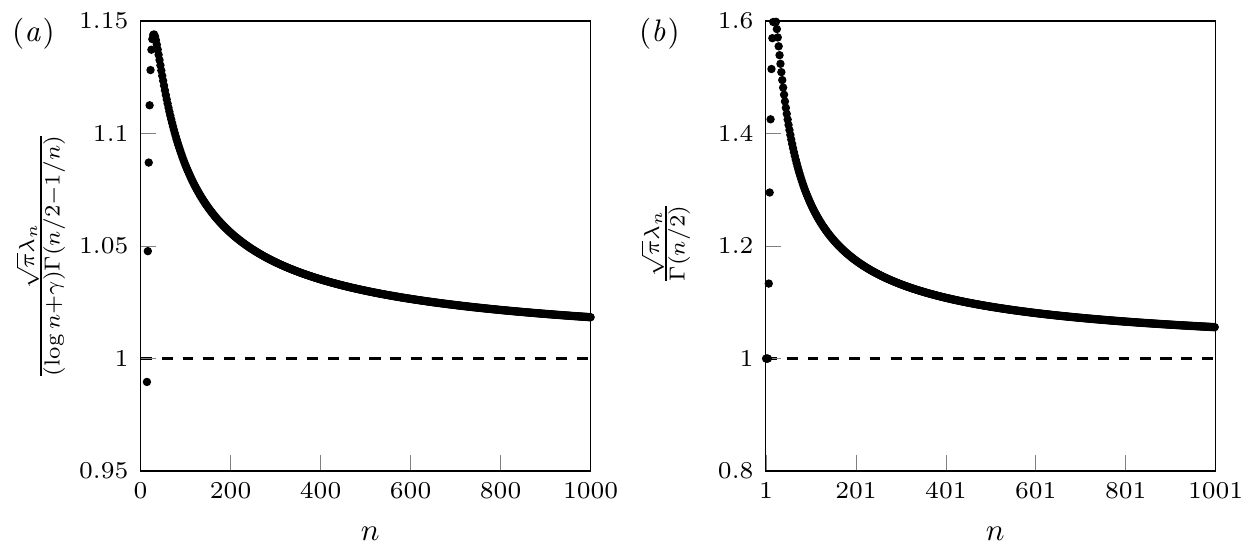}
	\caption{The coefficient $\lambda_n$, numerically calculated by the scheme
          of \S\ref{sec:jonYzero} is compared to the asymptotic prediction
          \eqref{eq:LateEig}. Comparison occurs for even $n$ in ($a$), and odd $n$ in ($b$).}%
	\label{fig:EigScaling}     
\end{figure}

We  now match the inner limit of the outer solution
\eqref{eq:Innertotalboth} with the outer-limit of the inner 
solution as  $\bar{y} \to -\infty$ given by \eqref{eq:InnerInf2}. Firstly,
since the inner solution 
contains no terms of $O(\log{n})$ for $n$ odd and $O(\log^2{(n)})$ for
$n$ even, we require 
\begin{equation}\label{eq:Matching1}
A_1 = -\frac{\delta_{2}}{4} \qquad \text{and} \qquad B_0 = -\frac{\delta_{0}}{4}.
\end{equation}
Next, matching each of the coefficients of $\e^{\bar{y}^2/2}/\bar{y}$ as $\bar{y} \to -\infty$ yields 
\begin{subequations}\label{eq:Matchingleftright}
\begin{equation}\label{eq:Matching2left}
 \bar{A}_{{L}}-\delta_{0}\sqrt{\frac{\pi}{8}}= -\frac{1}{\sqrt{2}}, \qquad
  \bar{A}_{{Q}}-\delta_{1}\sqrt{\frac{\pi}{8}}= -\frac{\gamma}{\sqrt{2}}, \qquad
\bar{A}_{{R}}-\delta_{2}\sqrt{\frac{\pi}{8}}= -\frac{1}{\sqrt{2}}.
\end{equation}
As $\bar{y} \to \infty$ we need the coefficients of
$\e^{\bar{y}^2/2}/\bar{y}$ to be zero, in order that the na\"ive
divergence is not present near the phantom singularity at $Y = +1$
(resolving issue 2 of \S\ref{sec:homo}).
Thus matching as $\bar{y} \to \infty$ gives
\begin{equation}\label{eq:Matching2right}
\bar{A}_{{L}}+\delta_{0}\sqrt{\frac{\pi}{8}}=0, \qquad 
\bar{A}_{{Q}}+\delta_{1}\sqrt{\frac{\pi}{8}}=0, \qquad 
\bar{A}_{{R}}+\delta_{2}\sqrt{\frac{\pi}{8}}=0.
\end{equation}
\end{subequations}
Solving \eqref{eq:Matching1}, \eqref{eq:Matchingleftright} gives
\begin{equation}\label{eq:Matching5}
\delta_{0} = \delta_{2} = \frac{1}{\sqrt{\pi}}, \qquad \delta_{1} =
\frac{\gamma}{\sqrt{\pi}},\qquad
\bar{A}_{{L}}  =  \bar{A}_{{R}}  = -
\frac{1}{\sqrt{8}}, \qquad  
\bar{A}_{{Q}}  =  - \frac{\gamma}{\sqrt{8}},
\end{equation}
which is consistent with $\delta_0=1/\sqrt{\pi}$ from \eqref{lanodd}.
In figure \ref{fig:EigScaling} we compare the asymptotic behaviour
\eqref{eq:LateEig} with $\la_n$ determined numerically following the procedure
described in \S\ref{sec:jonYzero}; the agreement validates our predictions for
$\delta_0$, $\delta_1$ and $\delta_2$.
      
\section{Stokes smoothing and determination of $\text{Im}[\lambda]$} \label{sec:smoothing}  
\begin{figure}[htpb]
	\centering
	\includegraphics[]{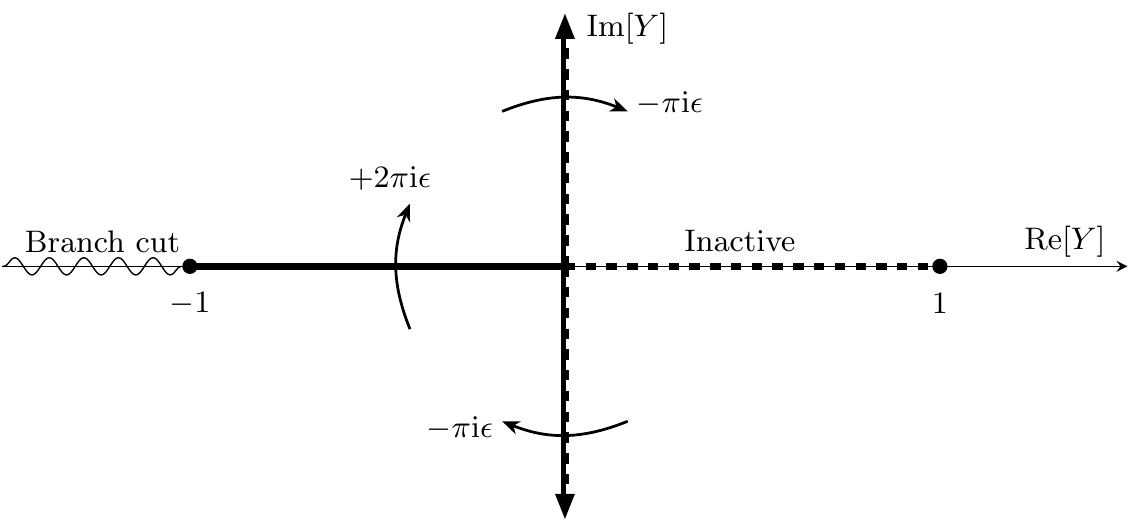}
	\caption{The Stokes lines generated by the divergent series expansion for our problem are shown (bold). Inactive Stokes lines are shown dashed, and along the imaginary axis the Stokes line has a multiplier of half the usual value. This inactivity is caused by the higher-order Stokes phenomenon, which switches off the na\"ive divergence across the imaginary axis.}%
	\label{fig:StokesLinesNew}     
\end{figure}

Having determined the form of the late terms we now  truncate the
divergent expansions for the solution and eigenvalue after $N$ terms and
study  the  remainder, by writing 
\begin{equation}
\label{eq:expremainder}
\psi(Y) = \underbrace{\sum _{n=0}^{N-1} \epsilon^n \psi_n(Y)}_{\psi_{\text{reg}}(Y)}+\mathcal{R}_N(Y) \quad \text{and} \quad \lambda= \underbrace{\sum_{n=0}^{N-1} \epsilon^n \lambda_n}_{\lambda_{\text{reg}}}+\laexp,
\end{equation}
where the truncated series are denoted by $\psi_{\text{reg}}(Y)$ and
$\lambda_{\text{reg}}$. We truncate optimally  by setting
\begin{equation}
\label{eq:optimalN}
N = \frac{2 \lvert \chi \rvert}{\ep^2}+\rho,
\end{equation}
where $0 \leq \rho <1$ ensures that $N$ takes integer values.
Substituting  into \eqref{eq:hermite} gives 
\begin{equation}
\label{eq:expremaindereq}
\ep^2 \mathcal{R}^{\prime \prime}_N - 2Y \mathcal{R}_N^{\prime}+ \bigg[\frac{\ep}{1+Y}-(1+\lambda_{\text{reg}}) \bigg]\mathcal{R}_N =\psi_{\text{reg}} \laexp+ \xi_{\text{eq}} + O(\laexp\mathcal{R}_N),
\end{equation}
where the forcing term $\xi_{\text{eq}}$ is of $O(\ep^N)$ and is defined by
\begin{equation}
\label{eq:xieq}
\xi_{\text{eq}}=(1+\lambda_{\text{reg}})\psi_{\text{reg}} - \ep^2 \psi_{\text{reg}}''+2Y\psi_{\text{reg}}'-\ep\frac{\psi_{\text{reg}}}{1+Y}. 
\end{equation}
As $\eps \ra 0$,
\begin{equation}
  \xi_{\text{eq}} \sim - \eps^{N+2} \psi_{N}''- \eps^{N+3}
  \psi_{N+1}''+\cdots.\label{remrhs}
\end{equation}
The procedure now is to:
\begin{enumerate}
\item Expand (\ref{remrhs}) as $\eps \ra
0$, $N \ra \infty$ using \eqref{eq:optimalN}; 
\item Write $\mathcal{R}_N$ as a Stokes multiplier $\mathcal{S}(Y)$ multiplied by a homogeneous
 solution by setting, in this case,
 $\mathcal{R}_N = \mathcal{S}(Y) \psi_{\mathrm{exp}}$
 with
 \begin{equation*}
 \psi_{\mathrm{exp}} =    \left( -\frac{1}{2Y}-\frac{\ep}{2Y}\left[
    \log (2/\eps^2) +  \gamma+\frac{\log{(1+Y)}}{2} \right]+\cdots\right) \e^{-(1-Y^2)/\ep^2};\label{eq:homogenousexp}
   \end{equation*}
\item Localise in a boundary layer near the Stokes lines where $\chi=1-Y^2$
is real and positive;
\item Solve for $\mathcal{S}$ to explicitly observe the rapid jump across
the Stokes line.
\end{enumerate}
Since these steps are fairly standard (see e.g. \cite{chapman_1998})
we omit the details here; the interested reader may refer to the geophysical study for the
Kelvin wave problem \cite{shelton2022Kelvin} where more details are given.
The upshot is that there is a jump in $\mathcal{S}$ of $2 \pi \i\ep$ as the
Stokes line $-1 \leq Y<0$ is crossed, so that a
multiple of $\psi_{\mathrm{exp}} $ is turned on, as shown  in figure~\ref{fig:StokesLinesNew}.

  While $\chi=1-Y^2$ is also real and positive on the imaginary axis,
   the Stokes line there is coincident with  the
  higher-order Stokes line across which the relevant contribution to
  $\psi_n$, including the right-hand side of
  \eqref{eq:expremaindereq}, is  switched
  off. The upshot is that the Stokes multiplier is multiplied by 1/2
  on this segment of the Stokes line. Finally, on the strip  $0 < Y<1$
  the relevant terms in the expansion of $\psi_n$ are no longer present,
  having been turned off by the higher-order Stokes phenomenon, so
  that this prospective Stokes line is inactive and no switching occurs.
  
The additional term $\psi_{\mathrm{exp}}$ switched on across the
  Stokes lines does not satisfy the decay condition as $Y \ra \infty$.
This term is cancelled by an additional contribution to
$\mathcal{R}_N$ generated by the forcing term $\psi_{\text{reg}}
\laexp$ due to the exponentially small correction to the
eigenvalue; indeed it is this requirement of cancellation which determines $\laexp$.
This additional particular solution satisfies (to two orders in $\ep$)
\begin{equation}
\label{eq:particular}
\epsilon^2 \mathcal{R}_N^{\prime \prime}-2Y\mathcal{R}_N^{\prime} \sim
\laexp,
\end{equation}
which may be solved in terms of special functions to find 
\begin{equation}
\label{eq:particularsol}
\mathcal{R}_N \sim \frac{\ep \laexp\sqrt{\pi}}{2Y} \e^{Y^2/\ep^2}
\end{equation}
as $Y \ra \infty$.

We now determine $\laexp$ by imposing that the coefficient of
$\e^{Y^2/\eps^2}/Y$ as $Y\ra\infty$ is zero.
First we note that the decay condition as $Y\ra-\infty$ may be
enforced on different Riemann sheets generated by the singularity at
$Y=-1$; essentially as we move from $Y=-\infty$ to $Y= + \infty$ we
  have to decide whether we pass above or below the point $Y=-1$.
If we pass above it the Stokes switching associated with the base
expansion
gives $-\pi \i \ep \psi_{\mathrm{exp}}$ at $Y = \infty$, while if we
pass below it gives $\pi \i \ep \psi_{\mathrm{exp}}$. This must cancel with the contribution from \eqref{eq:particularsol}, which gives
\begin{equation}\label{eq:eigenvalueprediciton}
\laexp \sim \pm \sqrt{\pi}\, \i \left[ 1-2 \ep \log \ep
  +\left(\gamma+\log 2\right) \ep\right] \e^{-1/\ep^2}.
\end{equation}
These are the complex-conjugate pairs for $\text{Im}[\lambda]$, which correspond to growing and decaying temporal instabilities in the solution.
We note that \eqref{eq:eigenvalueprediciton} is consistent with a
direct application of Borel summation to the divergent series
(\ref{eq:LateEig}), as we would expect.

\subsection{Conclusion and discussion} \label{sec:conclusion}
We have derived the exponentially-small component of the eigenvalue,
\begin{equation}
\text{Im}[\lambda] \sim \pm \sqrt{\pi}\Big[ 1-2 \ep \log \ep
+\left(\gamma+\log 2\right) \ep\Big] \e^{-1/\ep^2},\label{imla}
\end{equation}
by considering the Stokes phenomenon displayed by the solution,
$\psi(Y)$, throughout the complex plane. Since this
exponentially-small component of $\lambda$ is imaginary, it
corresponds to a growing temporal instability of the solution
associated with weak shear, and is known as a critical layer instability.

As we noted in \S\ref{sec:roadmap}, the Hermite-with-pole problem,
posed by \cite{boyd1998sturm} as a model for weak latitudinal shear of
the equatorial Kelvin wave, is an unusually difficult problem in
exponential asymptotics.

Some of the issues we have had to confront, such as the differing
asymptotic behaviours for even and odd terms in the expansion, arise
from an unfortunate choice of model equation, forcing the expansion to
proceed in powers of $\eps$ when it would more naturally proceed in
powers of $\eps^2$. 
Some, such as the divergence of the asymptotic
series for the eigenvalue, and its associated exponentially small
imaginary component, are more generic.

The logarithmic factors of $n$ in the behaviour of the late terms are
associated with the logarithmic factors of $\eps$ in the expansion of
the imaginary part of the eigenvalue (\ref{imla}). It is not clear to
what extent we were just unlucky to have to confront these, although
we note that had we only wanted the leading term in (\ref{imla}) we
could have avoided most (but not all) of the logs by only considering
only the  dominant (i.e. odd $n$) terms in the expansions of $\psi$ and $\la$.

The most interesting aspect of the problem has been the phantom
singularity in the na\"ive expansion of the late terms, and its
resolution via a higher-order Stokes phenomenon driven by the
divergent eigenvalue expansion. At the moment it is not clear to us
whether this behaviour is unusual or generic, but we hope the analysis
we have presented will act as a road map for similar problems.
Although we only considered the higher-order Stokes line in the
vicinity of $Y=0$, it is possible to show that it extends along the
whole imaginary axis, and to smooth it in a similar manner to the
smoothing of regular Stokes lines; the interested reader is referred to \cite{shelton2022HOSP}.

\section*{Acknowledgments}
The authors would like to thank the Isaac Newton Institute for Mathematical Sciences for support and hospitality during the programme Applicable Resurgent Asymptotics when work on this paper was undertaken. This work was supported by EPSRC Grant Number EP/R014604/1. We also thank Dr. Stephen Griffiths (Leeds) for many useful discussions and for hosting a short research visit, funded by the UK Fluids Network, where this work was initialised. PHT gratefully acknowledges support from EPSRC Grant Number EP/V012479/1.

\appendix

\bibliographystyle{jfm}

\providecommand{\noopsort}[1]{}
\begin{thebibliography}{17}
\expandafter\ifx\csname natexlab\endcsname\relax\def\natexlab#1{#1}\fi
\def\au#1{#1} \def\ed#1{#1} \def\yr#1{#1}\def\at#1{#1}\def\jt#1{\textit{#1}}
  \def\bt#1{#1}\def\bvol#1{\textbf{#1}} \def\vol#1{#1} \def\pg#1{#1}
  \def\publ#1{#1}\def\arxiv#1{#1}\def\org#1{#1}\def\st#1{\textit{#1}}

\bibitem[Berry(1989)]{berry_1989}
{\sc \au{Berry, M.~V.}} \yr{1989}  \at{Uniform asymptotic smoothing of {S}tokes
  discontinuities}.  \jt{Proc. R. Soc. Lond.}  \bvol{A 422},  \pg{7--21}.

\bibitem[Body {\em et~al.\/}(2005)Body, King \& Tew]{body2005exponential}
{\sc \au{Body, G.~L.}, \au{King, J.~R.} \& \au{Tew, R.~H.}} \yr{2005}
  \at{Exponential asymptotics of a fifth-order partial differential equation}.
  \jt{Eur. J. Appl. Math.}  \bvol{16}~(5),  \pg{647--681}.

\bibitem[Boyd \& Natarov(1998)]{boyd1998sturm}
{\sc \au{Boyd, J.~P.} \& \au{Natarov, A.}} \yr{1998}  \at{A
  {S}turm--{L}iouville eigenproblem of the fourth kind: A critical latitude
  with equatorial trapping}.  \jt{Stud. Appl. Math.}  \bvol{101}~(4),
  \pg{433--455}.

\bibitem[Chapman(1999)]{chapman_1999}
{\sc \au{Chapman, S.~J.}} \yr{1999}  \at{On the role of {S}tokes lines in the
  selection of {S}affman-{T}aylor fingers with small surface tension}.
  \jt{Eur. J. Appl. Math.}  \bvol{10}~(6),  \pg{513--534}.

\bibitem[Chapman {\em et~al.\/}(1998)Chapman, King \& Adams]{chapman_1998}
{\sc \au{Chapman, S.~J.}, \au{King, J.~R.} \& \au{Adams, K.~L.}} \yr{1998}
  \at{Exponential asymptotics and {S}tokes lines in nonlinear ordinary
  differential equations}.  \jt{Proc. R. Soc. Lond. A}  \bvol{454},
  \pg{2733--2755}.

\bibitem[Chapman \& Kozyreff(2009)]{chapman2009exponential}
{\sc \au{Chapman, S.~J.} \& \au{Kozyreff, G.}} \yr{2009}  \at{Exponential
  asymptotics of localised patterns and snaking bifurcation diagrams}.
  \jt{Phys. D}  \bvol{238}~(3),  \pg{319--354}.

\bibitem[Chapman \& Mortimer(2005)]{chapman_2005}
{\sc \au{Chapman, S.~J.} \& \au{Mortimer, D.~B.}} \yr{2005}  \at{Exponential
  asymptotics and {S}tokes lines in a partial differential equation}.
  \jt{Proc. R. Soc. Lond. A}  \bvol{461}~(2060),  \pg{2385--2421}.

\bibitem[Chapman \& Vanden-Broeck(2002)]{chapman_2002}
{\sc \au{Chapman, S.~J.} \& \au{Vanden-Broeck, J.-M.}} \yr{2002}
  \at{Exponential asymptotics and capillary waves}.  \jt{SIAM J. Appl. Math.}
  \bvol{62 (6)},  \pg{1872--1898}.

\bibitem[Daalhuis(2004)]{daalhuis2004higher}
{\sc \au{Daalhuis, A. B.~Olde}} \yr{2004}  \at{On higher-order {S}tokes
  phenomena of an inhomogeneous linear ordinary differential equation}.  \jt{J.
  Comp. Appl. Math.}  \bvol{169}~(1),  \pg{235--246}.

\bibitem[Dingle(1973)]{dingle_book}
{\sc \au{Dingle, R.~B.}} \yr{1973} {\em Asymptotic Expansions: Their Derivation
  and Interpretation\/}.  \publ{Academic Press, London}.

\bibitem[Howls {\em et~al.\/}(2004)Howls, Langman \& Daalhuis]{howls_2004}
{\sc \au{Howls, C.~J.}, \au{Langman, P.~J.} \& \au{Daalhuis, A. B.~Olde}}
  \yr{2004}  \at{On the higher-order {S}tokes {P}henomenon}.  \jt{Proc. R. Soc.
  Lond. A}  \bvol{460},  \pg{2285--2303}.

\bibitem[Kruskal \& Segur(1991)]{kruskal_1991}
{\sc \au{Kruskal, M.~D.} \& \au{Segur, H.}} \yr{1991}  \at{Asymptotics beyond
  all orders in a model of crystal growth}.  \jt{Stud. Appl. Math.}  \bvol{85},
   \pg{129--181}.

\bibitem[Natarov \& Boyd(2001)]{natarov2001beyond}
{\sc \au{Natarov, A.} \& \au{Boyd, J.~P.}} \yr{2001}  \at{Beyond-all-orders
  instability in the equatorial {K}elvin wave}.  \jt{Dynam. Atmos. Oceans}
  \bvol{33}~(3),  \pg{191--200}.

\bibitem[Shelton {\em et~al.\/}(2023{\natexlab{{\em a\/}}})Shelton, Chapman,
  Griffiths \& Trinh]{shelton2022Kelvin}
{\sc \au{Shelton, J.}, \au{Chapman, S.~J.}, \au{Griffiths, S.} \& \au{Trinh,
  P.~H.}} \yr{2023{\natexlab{{\em a\/}}}}  \at{On the exponentially-small
  instability of the equatorial {K}elvin wave}.  \jt{In Preperation} .

\bibitem[Shelton {\em et~al.\/}(2023{\natexlab{{\em b\/}}})Shelton, Crew \&
  Trinh]{shelton2022HOSP}
{\sc \au{Shelton, J.}, \au{Crew, S.} \& \au{Trinh, P.~H.}}
  \yr{2023{\natexlab{{\em b\/}}}}  \at{Exponential asymptotics and higher-order
  {S}tokes phenomenon in singularly perturbed {ODE}s}.  \jt{In Preperation} .

\bibitem[Shelton \& Trinh(2022)]{shelton2022exponential}
{\sc \au{Shelton, J.} \& \au{Trinh, P.~H.}} \yr{2022}  \at{Exponential
  asymptotics for steady parasitic capillary ripples on steep gravity waves}.
  \jt{J. Fluid Mech.}  \bvol{939},  \pg{A17}.

\bibitem[Tanveer(1987)]{tanveer1987analytic}
{\sc \au{Tanveer, S.}} \yr{1987}  \at{Analytic theory for the selection of a
  symmetric {S}affman--{T}aylor finger in a {H}ele--{S}haw cell}.  \jt{Phys.
  Fluids}  \bvol{30}~(6),  \pg{1589--1605}.

\end{thebibliography}
\providecommand{\noopsort}[1]{}

\section{Lower-order divergence of the na\"ive ansatz} \label{sec:LowerOrderNaive}
As noted in \S\ref{sec:homo}, the na\"ive factorial-over-power solution to the homogeneous late term equation \eqref{eq:latehom} is unable to satisfy the boundary condition at $Y=0$. This is due to a singularity in the prefactors of the divergent ansatz, $L(Y)$, $R(Y)$, and $Q(Y)$, from equations \eqref{eq:prefactor} and \eqref{eq:prefactorQ}. One may consider lower order terms, as $n \to \infty$, in the divergence of the homogeneous solution by considering a prefactor of the form (for $n$ odd)
\begin{equation}
\label{eq:prefactorexp}
  R(Y) = R_0(Y)+\frac{\log{(n)}}{n}M_1(Y)+\frac{R_1(Y)}{n} + \cdots,
\end{equation}
where the subsequent terms in this series will be of $O(n^{-2}\log{n})$ and $O(n^{-2})$.
We will see that the strength of the singularity in $R_0(Y)$ at $Y=0$ increases in later orders and thus forces a reordering of the series as $Y \to 0$. 

The method to calculate these lower order solutions is similar to that briefly presented in \S\ref{sec:homo} for the leading orders. We substitute an ansatz for $\psi_n(Y)$ of the form
\begin{equation} \label{eq:factoriallower}
\psi_{n} \sim
\left\{\begin{aligned}
 \bigg[L_0(Y)\log{(n)}+Q_0(Y) + \frac{\log{(n)}}{n}L_1(Y)+&\cdots \bigg] \frac{\Gamma(\frac{n}{2} -1)}{\chi^{n/2-1}} \quad \text{for $n$ even}, \\
 \bigg[R_0(Y) +\frac{\log{(n)}}{n}M_1(Y)+ \frac{R_1(Y)}{n} +&\cdots\bigg]\frac{\Gamma(\frac{n-1}{2} )}{\chi^{(n-1)/2}} \quad ~ \text{for $n$ odd},
\end{aligned}\right.
\end{equation}
into the homogeneous equation \eqref{eq:latehom}.
Dividing out by the dominant factorial-over-power scaling in the $O(\ep^n)$ equation \eqref{eq:latehom} yields terms of orders $n^0$, $n^{-1} \log{(n)}$, $n^{-1}$, $n^{-2}\log{(n)}$, and $n^{-2}$ for odd values of $n$. The case for even values of $n$ is similar, except for terms of order $\log{(n)}$ appearing. Distinct equations are found at each of these orders for the cases of $n$ even or $n$ odd.

The first few equations are the same as that considered in \S\ref{sec:homo}, and yield the singulant $\chi(Y)=1-Y^2$ from equation \eqref{eq:mychi} and prefactors $L_0(Y)$, $R_0(Y)$, and $Q_0(Y)$ from \eqref{eq:prefactoreq} and \eqref{eq:prefactorQ}. 
Equations for $M_1(Y)$ and $L_1(Y)$ are then found at $O(n^{-2}\log{(n)})$ for odd and even values of $n$, respectively, which have the solutions
\begin{equation}\label{eq:M1sol}
\begin{aligned}
M_1(Y)=&\bigg[ \Lambda_{M_{1}}+ \Lambda_{L_0} \log{(1+Y)}\bigg]\frac{(1-Y^2)}{Y},\\
L_1(Y)=& \bigg[ \Lambda_{L_1} + \frac{\Lambda_{M_1}}{2}\log{(1+Y)} +\frac{\Lambda_{L_0}}{Y^2}+\frac{\Lambda_{L_0}}{4}\log^2{(1+Y)}\bigg] \frac{(1-Y^2)}{Y},
\end{aligned}
\end{equation}
where $\Lambda_{M_1}$ and $\Lambda_{L_1}$ are constants of integration. It remains to determine $R_1(Y)$, the governing equation for which will be found at $O(n^{-2})$ when $n$ is odd. This has the solution 
\begin{equation}\label{eq:R1sol}
\begin{aligned}
R_1(Y)=&\bigg[\Lambda_{R_1}+\Lambda_{Q_0}\log{(1+Y)}+\frac{\Lambda_{R_0}}{Y^2}+\frac{\Lambda_{R_0}}{4} \log^2{(1+Y)}+\\
&-\Lambda_{M_1}\log{(1-Y^2)}-\Lambda_{L_0}\log{(1+Y)}\log^2{(1-Y^2)}\bigg]\frac{(1-Y^2)}{Y},
\end{aligned}
\end{equation}
where $\Lambda_{R_1}$ is a constant of integration.

To conclude, the functional prefactor of a factorial-over-power ansatz for the late-term solution may contain singularities or branch points at locations not seen in the early orders of the expansion. In our case this is the location $Y=0$. In these instances, it is necessary to consider lower order terms of the ansatz in order to determine the correct inner-variable scaling for the resultant boundary layer matching procedure.

\section{Inner solution at the singularity $Y=-1$}\label{sec:innersolnew}
Motivated by the inner limit of the outer solution, \eqref{eq:innerLim}, we consider an inner solution of the form 
\begin{equation}
\label{eq:innerSol}
\hat{\psi}_{\text{inner}}(\hat{y}) = \sum^{\infty}_{n=0} \sum^{n}_{m=0} \ep^n \log^m{(\ep)} \hat{\psi}_{(n,m)}(\hat{y}).
\end{equation}
Substitution into the inner equation \eqref{eq:innerEq} yields at $O(1)$, $O(\ep \log{(\ep)})$, and $O(\ep^2 \log^2{(\ep)})$
\begin{equation}
\label{eq:innerO1}
\widehat{\mathcal{L}}[\hat{\psi}_{(0,0)}] \equiv \dd{^2 \hat{\psi}_{(0,0)}}{\hat{y}^2} + 2\dd{\hat{\psi}_{(0,0)}}{\hat{y}}=0, \qquad \widehat{\mathcal{L}}[\hat{\psi}_{(1,1)}]=0,\qquad  \widehat{\mathcal{L}}[\hat{\psi}_{(2,2)}]=0.
\end{equation}
These equations have solutions of a similar form, given by $\hat{\psi}_{(0,0)}(\hat{y})=A_{(0,0)} + B_{(0,0)} \exp{(-2 \hat{y})}$ for instance. Matching with the $O(1)$, $O(\ep \log{(\ep)})$, and $O(\ep^2 \log^2{(\ep)})$ components of the inner-limit of $\psi_{\text{outer}}$ in \eqref{eq:innerLim} requires the coefficient of $\exp{(-2\hat{y})}$ to be zero for each of these solutions. Matching with the constant components then yields
\begin{equation}
\label{eq:innerO1sol}
\hat{\psi}_{(0,0)}(\hat{y}) = 1, \qquad \hat{\psi}_{(1,1)}(\hat{y}) = -1, \qquad \hat{\psi}_{(2,2)}(\hat{y}) = \frac{1}{2}.
\end{equation}
Next, at $O(\ep)$ and $O(\ep^2 \log{\ep})$, we find similar equations to \eqref{eq:innerO1} with the exception of a forcing term that relies on $\hat{\psi}_{(0,0)}(\hat{y})$ and $\hat{\psi}_{(1,1)}(\hat{y})$, respectively. These equations are found to be
\begin{equation}
\label{eq:innerO2}
\widehat{\mathcal{L}}[\hat{\psi}_{{(1,0)}}]=-\frac{1}{\hat{y}} \qquad \text{and} \qquad \widehat{\mathcal{L}}[\hat{\psi}_{(2,1)}]=\frac{1}{\hat{y}},
\end{equation}
where $\widehat{\mathcal{L}}$ is the linear differential operator defined in \eqref{eq:innerO1}.
For brevity, only the exact solution of the first of these is provided here. This has the solution of
\begin{equation}
\label{eq:innerO2sol}
\hat{\psi}_{(1,0)}(\hat{y}) = A_{(1,0)}+B_{(1,0)} \mathrm{e}^{-2\hat{y}} - \mathrm{e}^{-2 \hat{y}} \int_{0}^{\hat{y}} \log{(y)} \mathrm{e}^{2y} \mathrm{d} y.
\end{equation}
Analogously for the second equation in \eqref{eq:innerO2} the exact solution will have constants $A_{(2,1)}$ and $B_{(2,1)}$, and a positive sign (+) in front of the last component of the solution in \eqref{eq:innerO2sol}.
To facilitate matching with the $O(\ep)$ outer solution, we take the outer-limit of \eqref{eq:innerO2sol} as $\hat{y} \to \infty$, yielding
\begin{equation}\label{eq:innerO2solOutLim}
\hat{\psi}_{(1,0)}(\hat{y}) \sim -\frac{1}{2}\log{(\hat{y})} + \frac{1}{2} \sum_{k=1}^{\infty} \frac{\Gamma{(k)}}{(2\hat{y})^k} \quad \text{and} \quad
\hat{\psi}_{(2,1)}(\hat{y}) \sim \frac{1}{2}\log{(\hat{y})} - \frac{1}{2} \sum_{k=1}^{\infty} \frac{\Gamma{(k)}}{(2\hat{y})^k}.
\end{equation}
Here we set $A_{(1,0)}=0$, $B_{(1,0)}=0$, $A_{(2,1)}=0$, and $B_{(2,1)}=0$ to match with the $O(\ep)$ term of the inner limit of the outer solution from \eqref{eq:innerLim}.

Note that we have been able to construct an exact solution for $\hat{\psi}_{(1,0)}$ and $\hat{\psi}_{(2,1)}$. In general, and typically for nonlinear problems, this is not possible and an ansatz must be introduced to capture the series expansion of the outer-limit behaviour of $\hat{\psi}(\hat{y})$, from which the coefficients of this series, in our case $\Gamma{(k)}$, would determined via the solution to a recurrence relation problem. This will be the approach used when considering the $O(\ep^2)$ equation,
\begin{equation}
\label{eq:innerO3}
\widehat{\mathcal{L}}[\hat{\psi}_{(2,0)}]= \frac{\log{(\hat{y})}}{2\hat{y}} - \sum_{k=1}^{\infty} \frac{\Gamma{(k)}}{(2\hat{y})^{k+1}},
\end{equation}
for which we consider a series expansion as $\hat{y} \to \infty$ of the form
\begin{equation}
\label{eq:innerO3series}
\hat{\psi}_{(2,0)}(\hat{y}) = \frac{\log^2{(\hat{y})}}{8} + \log{(\hat{y})} \sum_{k=1}^{\infty} \frac{a_k}{(2 \hat{y})^{k}} + \sum_{k=1}^{\infty} \frac{b_k}{(2 \hat{y})^{k}}.
\end{equation}
Substitution of series \eqref{eq:innerO3series} into equation \eqref{eq:innerO3} yields terms that are either algebraic powers of $(2\hat{y})^{-k}$ or $\log{(y)} (2 \hat{y})^{-k}$. Examining the equations which arise at each of these orders yields the following recurrence relations for $a_k$ and $b_k$,
\begin{equation}
\label{eq:akeq}
\begin{aligned}
a_1=-\frac{1}{4},& \qquad a_k=(k-1)a_{k-1}, \\
b_1=\frac{1}{4},& \qquad  b_k=(k-1)b_{k-1}  + \frac{(2k-1)}{4k}\Gamma(k-1),
\end{aligned}
\end{equation}
where $k \geq 2$.
In substituting for $b_k = \Gamma(k) d_k$, the recurrence relation for $b_k$ may be written in a form with s series solution, yielding for $k \geq 2$
\begin{equation}
\label{eq:aksol}
a_k= -\frac{\Gamma(k)}{4} \qquad \text{and} \qquad  b_k  = \bigg[ \frac{1}{2}  - \frac{1}{4k} +\frac{1}{2}\sum_{j=2}^{k} \frac{1}{j}\bigg]\Gamma(k).
\end{equation}
Thus, as $\hat{y} \to \infty$, our $O(\ep^2)$ inner solution is given by
\begin{equation}
\label{eq:innerO3seriesfinal}
\hat{\psi}_{(2,0)}(\hat{y}) = \frac{\log^2{(\hat{y})}}{8} - \frac{\log{(\hat{y})}}{4} \sum_{k=1}^{\infty} \frac{\Gamma(k)}{(2 \hat{y})^{k}} + \sum_{k=1}^{\infty} \frac{b_k}{(2 \hat{y})^{k}},
\end{equation}
where $b_k$ is defined in equation \eqref{eq:aksol}. In \S{\ref{sec:innersolouter}}, we use the outer limit of this solution to motivate the correct form for the factorial-over-power ansatz of $\psi_n$ as $n \to \infty$. Thus, we are also interested in the limit of $k \to \infty$ of $b_k$. Expanding $b_k$ given in \eqref{eq:aksol} as $k \to \infty$ yields
\begin{equation}
\label{eq:bksollim}
 b_k  \sim  \bigg[\frac{1}{2} \log{(k)} + \frac{\gamma}{2} +O(k^{-1})\bigg] \Gamma(k),
\end{equation}
where $\gamma \approx 0.577$ is the Euler-Macheroni constant.

\end{document}